\documentclass[fleqn,usenatbib]{mnras}

\usepackage{newtxtext,newtxmath}
\usepackage[T1]{fontenc}

\DeclareRobustCommand{\VAN}[3]{#2}
\let\VANthebibliography\thebibliography
\def\thebibliography{\DeclareRobustCommand{\VAN}[3]{##3}\VANthebibliography}

\usepackage{graphicx}	% Including figure files
\usepackage{amsmath}	% Advanced maths commands
\usepackage{xcolor}
\usepackage{ulem}
\usepackage{soul}
\title[\textit{JWST} CEERS T-dwarf]{A T-Dwarf Candidate from \textit{JWST} Early Release NIRCam data}

\author[Wang et al. 2023]{
Po-Ya Wang$^{1,2}$
\thanks{E-mail: \href{mailto:poyawang@gapp.nthu.edu.tw}{poyawang@gapp.nthu.edu.tw}},
Tomotsugu Goto$^{1,2}$,
Simon C.-C. Ho$^{3}$,
Yu-Wei Lin$^{1,2}$,
Cossas K.-W. Wu$^{1,2}$,
\newauthor
Chih-Teng Ling$^{1}$,
Tetsuya Hashimoto$^{4}$,
Seong Jin Kim$^{1}$ and 
Tiger Y.-Y. Hsiao$^{5}$
\\
$^{1}$Institute of Astronomy, National Tsing Hua University, 101, Section 2. Kuang-Fu Road, Hsinchu, 30013, Taiwan (R.O.C.)\\
$^{2}$Department of Physics, National Tsing Hua University, 101, Section 2. Kuang-Fu Road, Hsinchu, 30013, Taiwan (R.O.C.)\\
$^{3}$Research School of Astronomy and Astrophysics, The Australian National University, Canberra, ACT 2611, Australia\\
$^{4}$Department of Physics, National Chung Hsing University, 145, Xingda Road, Taichung, 40227, Taiwan (R.O.C.)\\
$^{5}$Center for Astrophysical Sciences, Department of Physics and Astronomy, The Johns Hopkins University, 3400 N Charles St. Baltimore, MD 21218, USA
}

\date{Accepted 2023 June 01. Received 2023 May 31; in original form 2022 December 05}

\pubyear{2023}

\begin{document}
\label{firstpage}
\pagerange{\pageref{firstpage}--\pageref{lastpage}}
\maketitle

\begin{abstract}
We present a distant T-type brown dwarf candidate at $\approx2.55$\,kpc discovered in the Cosmic Evolution Early Release Science (CEERS) fields by \textit{James Webb Space Telescope (JWST)} NIRCam. In addition to the superb sensitivity, we utilised 7 filters from \textit{JWST} in near-IR and thus is advantageous in finding faint, previously unseen brown dwarfs. From the model spectra in new \textit{JWST}/NIRCam filter wavelengths, the selection criteria of F115W-F277W$<$-0.8 and F277W-F444W$>$1.1 were chosen to target the spectrum features of brown dwarfs having temperatures from 500\,K to 1300\,K. Searching through the data from Early Release Observations (ERO) and Early Release Science (ERS), we find 1 promising candidate in the CEERS field. The result of SED fitting suggested an early T spectral type with a low effective temperature of T$_\text{eff}\approx$1300\,K, the surface gravity of $\log{g}\approx5.25\,\text{cm s}^{-2}$, and an eddy diffusion parameter of logK$_{zz}\approx7\,\text{cm}^2 \text{s}^{-1}$, which indicates an age of $\approx$1.8\,Gyr and a mass of $\approx0.05$\,M$_{\odot}$. In contrast to typically found T dwarf within several hundred parsecs, the estimated distance of the source is $\approx2.55$\,kpc, showing the \textit{JWST}’s power to extend the search to a much larger distance.

\end{abstract}

\begin{keywords}
brown dwarfs -- infrared: stars -- stars: fundamental parameters
\end{keywords}

%%%%%%%%%%%%%%%%%%%%%%%%%%%%%%%%%%%%%%%%%%%%%%%%%%

%%%%%%%%%%%%%%%%% BODY OF PAPER %%%%%%%%%%%%%%%%%%

\section{Introduction}%_____________________________________
\label{sec:Introduction}
    It is important to understand brown dwarfs as they bridge the gap between the stellar and planetary mass regimes. However, because of their low temperatures, they are faint and difficult to be found. 
   % Brown dwarfs are sub-stellar objects with a mass between Jupiter and the main sequence stars, which are insufficient to trigger hydrogen fusion. 
    Previously, a census of L, T, and Y dwarfs was performed through \textit{Gaia} observations \citep{2021ApJS..253....7K}. With \textit{Spitzer} astrometry confirming the distance of these objects, a list of 525 sources complete out to 20\,pc was reported. Although a number of T and Y dwarfs were found through \textit{Wide-field Infrared Survey Explorer (WISE)/ Near-Earth Object Wide-field Infrared Survey Explorer (NEOWISE)}, the sensitivity restricted the search volume to the solar vicinity. 
    
    For T-type or below, with a very low effective temperature (below approximately 1300\,K), brown dwarfs are faint in optical and predicted to be brightest in near-IR \citep{2015ARA&A..53..279M}. To further distinguish the stellar type of these objects, the rich chemical absorption feature can be used.
    
    Previously both WISE/NEOWISE and Spitzer had only 4 filters in near-IR. In contrast,  \textit{James Webb Space Telescope (JWST)} NIRCam has 29 filters in the range of 0.6 to 5.0\,$\mu$m, significantly increasing efficiency in finding these cold objects within near-IR observations. Recently, \citet[]{2022arXiv220714802N} reported a possible galactic thick disk/halo brown dwarf candidate found in the JWST Early Release Science Abell 2744 parallel field (GLASS) using NIRCam. The candidate was discovered as late-T type with a low effective temperature of 650\,K, showing the power of \textit{JWST} to reveal these cool, distant objects.  
    
    We present a search of brown dwarfs through \textit{JWST} Early Release Observations (ERO) and Early Release Science (ERS) in this work, and a discovery of a possible T dwarf candidate. In Section \ref{sec:Methods}, we summarise the observation and candidate selection. In Section \ref{sec:Analysis}, we discuss the physical properties of the candidate evaluated from the Spectral Energy Distribution (SED) fitting result and the expected number of the searched fields. In Section \ref{sec:conclusion}, we summarise the result of this work. AB magnitude system is used throughout the paper.

\section{Methods}%____________________________________________
\label{sec:Methods}
    \subsection{Observations}
    \label{sec:obseravtion}
        This work utilised four observations from ERO and ERS with their availability of NIRCam. We summarise the observations as follows.
        
        The Cosmic Evolution Early Release Science (CEERS) Survey (PID: 1345) observed the Extended Groth Strip HST legacy field in June 2022 using \textit{JWST} NIRCam and MIRI with follow-up NIRSpec and other fields of MIRI in December 2022. In this work, we used the first four NIRCam pointings: CEERS1, CEERS2, CEERS3 and CEERS6; obtaining 7 filters: F115W, F150W, F200W, F277W, F356W, F444W, F410M.
        
        The \textit{JWST} Early Release Observation 6 (PID: 2732) observed the Stephan's Quintet compact group (hereafter SQ), with 3 MIRI filters and 6 NIRCam filters imaging. We used these NIRCam filters: F090W, F150W, F200W, F277W, F356W and F444W.
        
        The \textit{JWST} Early Release Observation 10 (PID: 2736) observed the galaxy lensing cluster SMACS-J0723.3-7327 (hereafter SMACS0723) with 6 NIRCam filters and 4 MIRI filters imaging. F090W, F150W, F200W, F277W, F356W and F444W were used in this work.
        
        Through the looking GLASS: a \textit{JWST} exploration of galaxy formation and evolution from cosmic dawn to present day (PID: 1324, hereafter GLASS) observed the lensing cluster Abell 2744 with NIRISS, NIRSpec and NIRCam parallel imaging. All 7 NIRCam filters are used in this work: F090W, F115W, F150W, F200W, F277W, F356W and F444W.
         
        We adopted the \textit{JWST} official observation source catalogue and images for CEERS, SQ, and SMACS fields. We analysed the GLASS field from the data release provided by \citet[]{2022ApJ...938L..14M}.
        
    \subsection{Candidate Selection}
    \label{sec:data_selection}

\begin{figure*}
        	\includegraphics[width=0.8\textwidth]{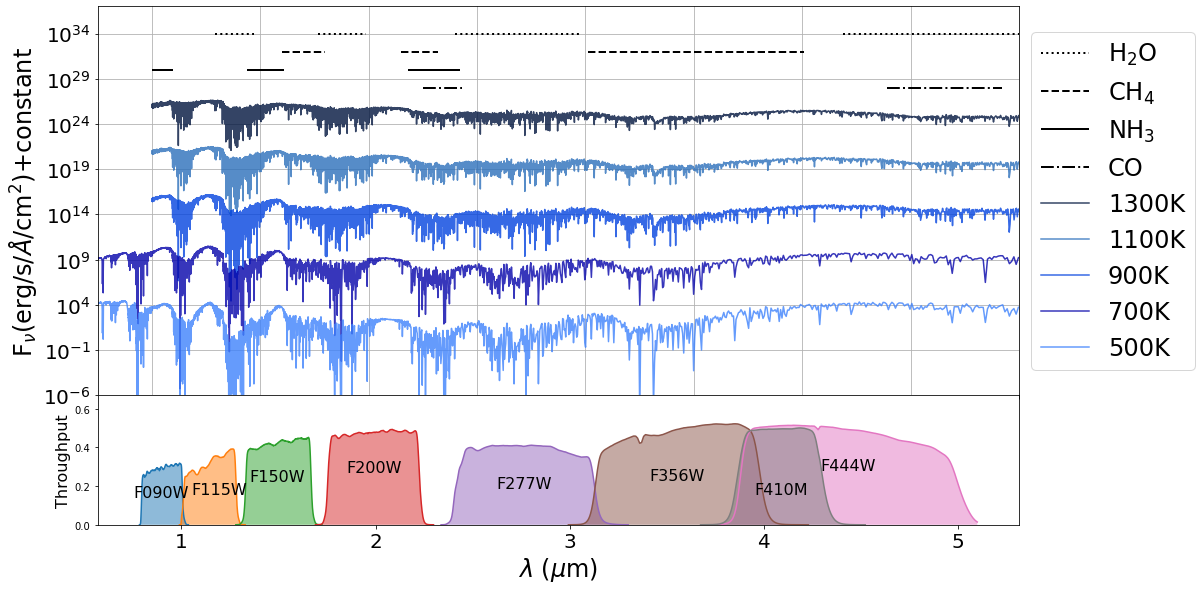}
            \caption{Model spectra with a fixed surface gravity of $g=10^{4.5}\,(\text{cm s}^{-2})$ and eddy diffusion coefficient of $K_{zz}=10^2\,(\text{cm}^{2} \text{s}^{-1})$, adopted from \citet[]{2021ApJ...923..269K}. We also show molecular absorption bands and the throughputs of \textit{JWST} NIRCam filters used in this work.}
            \label{fig:SCspectra}
        \end{figure*}

        We chose the brown dwarf model, Sonora-Cholla \citep{2021ApJ...923..269K,2021ApJ...920...85M}, as they provided detailed spectra modelling for near-IR wavelength. The chemical disequilibrium assumptions utilised in the model were also described to be more accurate for the atmosphere of low-temperature stellar objects \citep{2016SSRv..205..285M}. Taking advantage of the numerous filters available in NIRCam observations, we defined specific colour-colour criteria to identify potential brown dwarf candidates. As Figure \ref{fig:SCspectra} shows, an absorption feature of water molecules appears in the wavelength range of F277W, while F115W and F444W remain relatively bright across all the temperatures. Therefore, we decided to apply colour-colour selection criteria of F115W$-$F277W$<$-0.8$\cap$F277W$-$F444W$>$1.1. For observations without F115W (e.g., SQ, SMACS), we used F277W-F444W$>$1.1$\cap$-0.3$>$F150W-F277W$>$-1.5.
        
\begin{figure}
        	\includegraphics[width=\columnwidth]{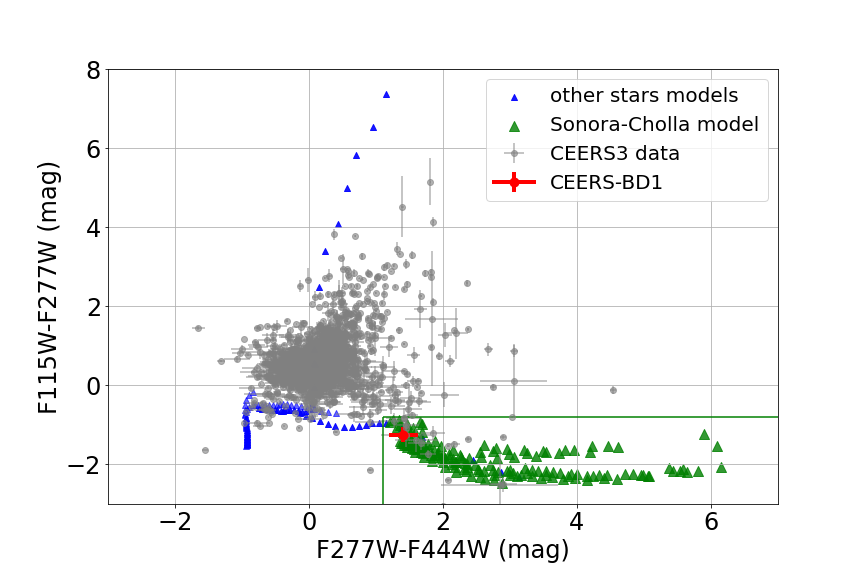}
            \caption{F115W-F277W-F444W colour-colour plot. Grey dots are observation data from \textit{JWST} CEERS3 field, green triangles are brown dwarf model predictions from \citet[]{2021ApJ...923..269K}, and blue triangles are other main sequence stellar models from \citet[]{1998PASP..110..863P} and \citet[]{1991A&A...250..370B}. The green line bounded area shows the colour selection criteria, and the red point presents the brown dwarf candidate CEERS-BD1 reported in this work.}
            \label{fig:ccplot}
        \end{figure}

        Figure \ref{fig:ccplot} shows the colour-colour plot of the CEERS3 field, where we show the observed data points and two sets of predictions from the Sonora-Cholla model and other main sequence stellar models (e.g.,\citet[]{1998PASP..110..863P,1991A&A...250..370B}). It is clear that the green lines bounded region (presenting our criteria of F115W$-$F277W$<$-0.8$\cap$F277W$-$F444W$>$1.1) excluded other main sequence stellar models, which ensure a well-fitted source in the region is possibly a brown dwarf. The red dot is the brown dwarf candidate we selected by the following descriptions, we further discussed the properties of the candidate in Section \ref{sec:Analysis_Physical_Properties}. 
        
        %After colour-colour selection, we filtered 260 sources out of the total of 26348 sources in all the fields and performed SED fitting with \textsc{LePhare} for these sources. 
        
        In order to exclude galaxies in the selection, we ran \textsc{SExtractor} using images in each field and tried several different parameters from both \textsc{SExtractor} result and official catalogue, e.g., FWHM, CLASS\_STAR, is\_extended and ELLIPTICITY. By eye-balling images, we found CLASS\_STAR performed best to filter out extended sources. Therefore, we selected sources with CLASS\_STAR$>$0.9. We have also masked out the edge of images and the foreground sources (e.g., Stephan’s Quintet in the SQ field, Abell 2744 in the GLASS field and SMACS J0723). Here, we filtered 1809 sources out of a total of 26348 sources, then performed the colour-colour selection mentioned above. 14 sources are left after the selection, but we found 13 of them are saturated after checking the images, therefore we excluded these. The only remained source is located in the CEERS3 field, with CLASS\_STAR=0.97.
        
        Note that the CEERS field is located in the well-studied Extended Groth Strip (EGS) field, therefore we also cross-matched to the CANDELS/EGS catalogue \citep{2017ApJS..229...32S} but found this source is not detected by any of the previous observations. Because its F150W=27.88$\pm$0.05 mag is slightly fainter than the 5 $\sigma$ depth of CANDELS/EGS catalogue detection band, \textit{HST}/WFC3 F160W=27.6 mag, it was excluded.

        We then performed a SED fitting using \textsc{LePhare} \citep{2006A&A...457..841I,1999MNRAS.310..540A} with Sonora-Cholla templates and found the source well fitted to an effective temperature T$_\text{eff}\approx$1300\,K model, where the three free parameters and their ranges are T$_\text{eff}\in[500K,1300K] $, surface gravity $\log{g}\in[3.5,5.5]$ and eddy diffusion parameter $\log{K_{zz}}\in[2,7]$.
        
\begin{figure}
            \includegraphics[width=\columnwidth]{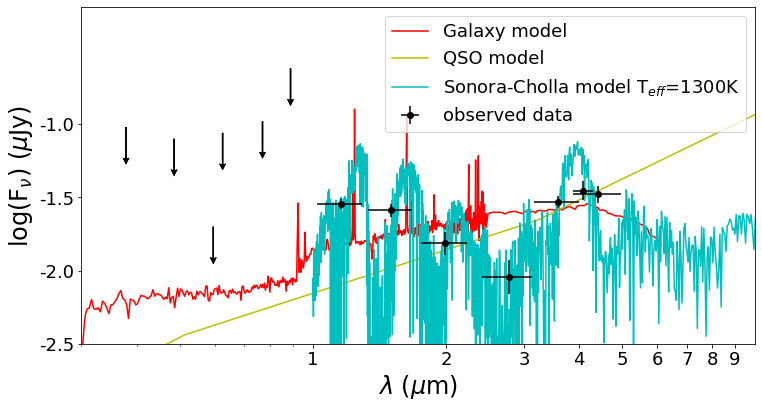}
            \caption{SED fitting result with \textit{Le PHARE}. The cyan line presents the best-fit stellar model of Sonora-Cholla \citep{2021ApJ...923..269K}, the red line presents the best-fit result from galaxy templates, and the green line presents the best-fit result from QSO templates. Both galaxy and QSO templates are originally included in \textsc{Le PHARE} \citep{2006A&A...457..841I,1999MNRAS.310..540A}. The upper limits are non-detection from \textit{CFHT}/Megacam u*, g', r', i', z' bands, and \textit{HST} ACS/F606W.}
            \label{fig:SED}
        \end{figure}

        Figure \ref{fig:SED} shows the SED fitting result with NIRCam observed photometric data and upper limits from \textit{CFHT}/Megacam u*, g', r', i', z' bands and \textit{HST} ACS/F606W observations. Also shown are three best-fitted results (determined by having smallest reduced $\chi^2$ value) from Sonora-Cholla, galaxy, and quasar (QSO) templates having reduced $\chi^2$ of 11.9, 54.8 and 25.1 respectively. 
        From the \textsc{Le PHARE} SED library, we adopted the CWW\_Kinney spectra for galaxies \citep{1980ApJS...43..393C, 1994ApJ...429..582C} and all QSO spectra from various authors which included observed and synthetic spectra \citep{2008MNRAS.386..697R,2007ApJ...666..806N,1998ApJ...509..103S}. For stellar SEDs, in addition to the library in the Le PHARE \citep{1998PASP..110..863P,2000ApJ...542..464C,1994PASP..106..566H}, we manually added Sonora-Cholla \citep{2021ApJ...923..269K}. 
        
        By comparison to these results, the object is more likely to be a star, rather than a compact galaxy or a distant QSO. The object is possibly a brown dwarf candidate, hereafter CEERS-BD1. 
        
        \begin{table}
        	\centering
        	\caption{Properties of CEERS-BD1.}
        	\label{tab:BD_info}
        	\begin{tabular}{lcr}
        		\hline
    		    R.A. & 14:19:14.61\\
    		    Dec & 52:53:00.19\\
                    \hline
                    &\textit{Le PHARE}& MCMC\\
                    \hline                    
    		    T$_{\text{eff}}$\,(K) & $1300$ & $1295\pm5$\\[5pt]
    		    
                    $\log{g}$\,$(\text{cm s}^{-2})$ & $5.25$ & $5.2^{+0.2}_{-0.5}$\\[5pt]
    		    
    		    $\log{K_{\text{zz}}}$$\,(\text{cm}^2 \text{s}^{-1})$ &  $7$ & $5.4\pm1.1$\\
    		    
                    \hline
                    
    		    d\,(kpc)  & $2.55^{+0.33}_{-0.48}$\\[5pt]
    		    
    		    mass\,(M$_{\odot}$) & $0.052^{+0.015}_{-0.022}$\\[5pt]
          
                    R\,(R$_{\odot}$) & $0.089^{+0.016}_{-0.011}$\\[5pt]
                
                    age\,(Gyr) & $1.8^{+6.4}_{-1.2}$\\
        	
        		\hline
        		band & AB mag\\
        		\hline
                F115W & $27.77\pm0.04$\\
                F150W & $27.88\pm0.05$\\
                F200W & $28.45\pm0.10$\\
                F277W & $29.02\pm0.14$\\
                F356W & $27.75\pm0.05$\\
                F410M & $27.55\pm0.08$\\
                F444W & $27.61\pm0.07$\\
        		\hline
        	\end{tabular}
        \end{table}
        
        \begin{figure}
        	\includegraphics[width=\columnwidth]{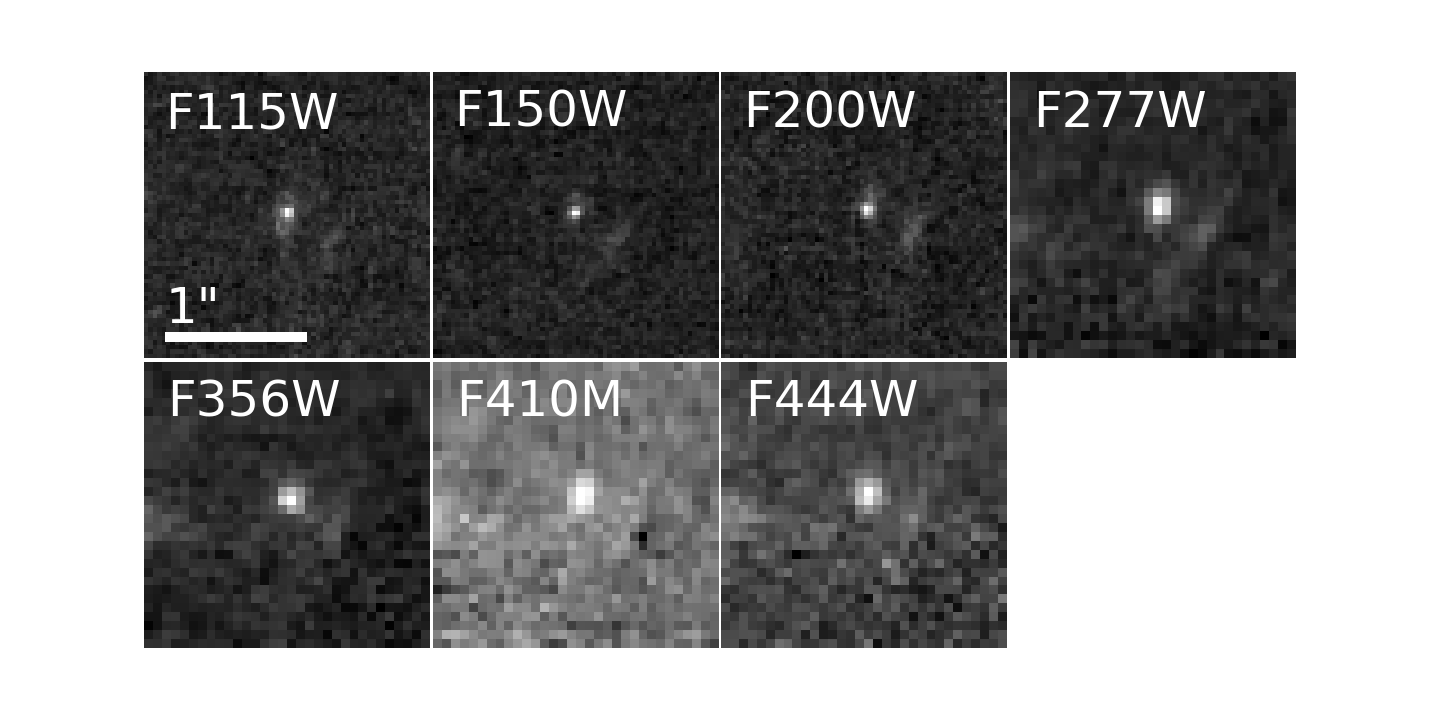}
            \caption{The \textit{JWST} NIRCam $2" \times 2"$ image cutout of CEERS-BD1.}
            %TODO-update the image
            \label{fig:images}
        \end{figure}
        
\section{Analysis and Discussion}%__________________________________________
\label{sec:Analysis}
    
    \subsection{Physical Properties}
    \label{sec:Analysis_Physical_Properties}
    The photometric data and some further properties of CEERS-BD1 discussed in the following section are summarized in Table \ref{tab:BD_info}. Compared to the photometric table and the evolution table \citep{2021ApJ...920...85M}, the template values of the best-fitted model indicate that CEERS-BD1 has approximately a low effective temperature of T$_\text{eff}\approx$1300\,K, the surface gravity of $\log{g}\approx5.25\,\text{cm s}^{-2}$, and an eddy diffusion parameter of $\log{K_{zz}}\approx7$\,\text{cm}$^2 \text{s}^{-1}$. 

\begin{figure*}
            \includegraphics[width=0.6\textwidth]{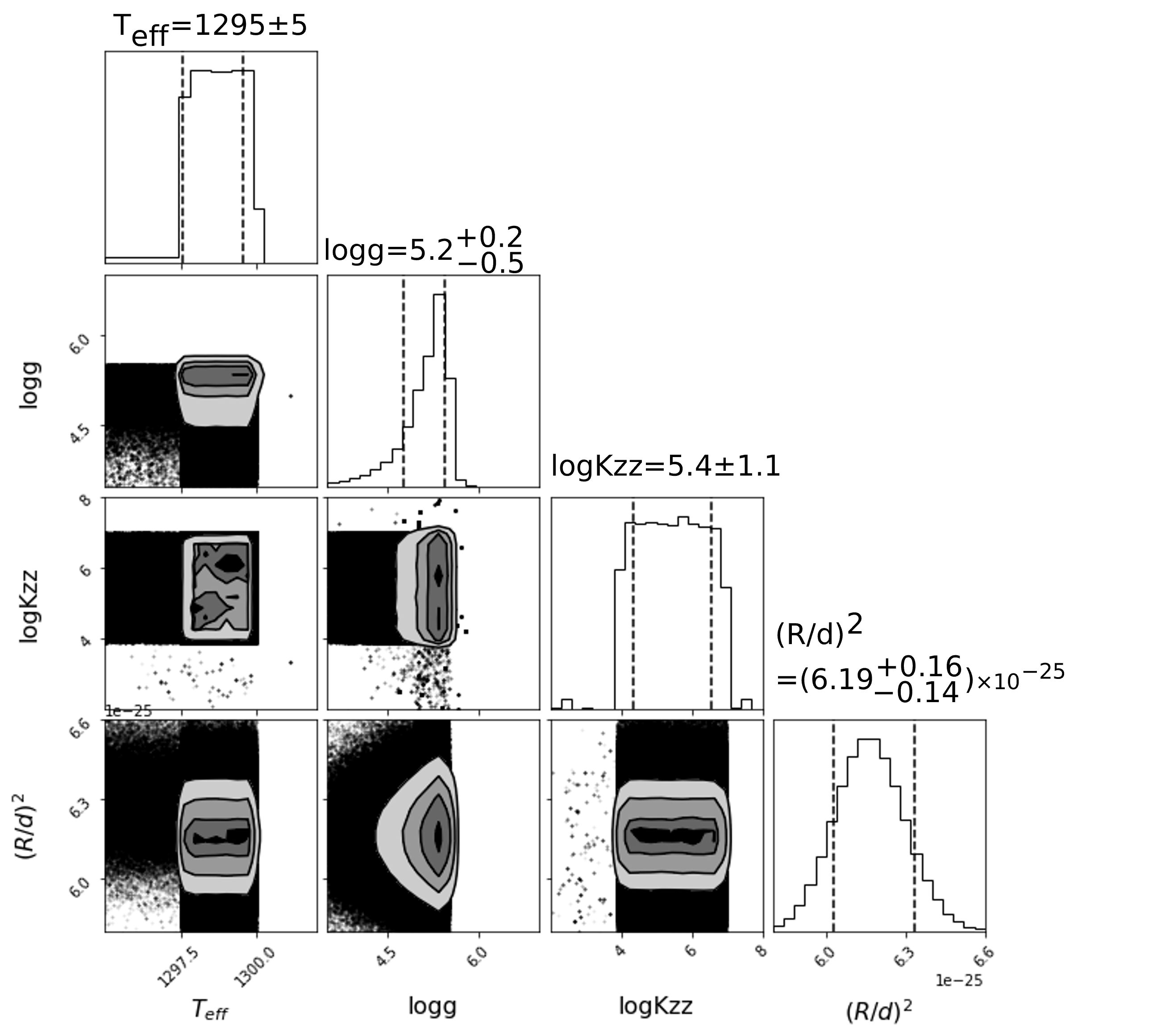}
            \caption{The posterior distribution of the MCMC for four physical parameters.}
            \label{fig:MCMC}
    \end{figure*}

    In order to further estimate the uncertainties of CEERS-BD1’s physical properties we performed Markov-Chain Monte Carlo (MCMC) chain using \textsc{Python} with package \textsc{emcee}. We linearly interpolated the Sonora-Cholla template in finer grids before performing the fitting. We use step sizes of $5$ K for T$_\text{eff}$, $0.1$ for $\log{g}  $ and $0.125$ for $\log{K_{zz}}$. The algorithm simply maximizes the likelihood function for the JWST NIRCam observed photometry to the template photometry which we derived through the NIRCam filter transmissions. A uniform prior was set within the parameter range of the Sonora-Cholla model and negative infinity probability for out of the range. Other than the three free parameters, we have the fourth parameter, the scaling factor of flux $(R/d)^2$, where $R$ presents the radius of the star and $d$ is the distance to the star. We then performed the MCMC chain with 200 walkers and 20,000 steps. The posterior distribution of the chains leads to a result of T$_\text{eff}=1295\pm5$, $\log(g)=5.2^{+0.2}_{-0.5}$, $\log{K_{zz}}=5.4\pm1.1$ and $(R/d)^2=6.19^{+0.16}_{-0.14}\times10^{-25}$. The posterior distribution of MCMC result is shown in Figure \ref{fig:MCMC}.

    We select the solar metallicity evolution table with fixed effective temperature and gravity \citep{2021ApJ...920...85M} in order to investigate further properties of the source. From template values of effective temperature and gravity, an age of $1.8^{+6.4}_{-1.2}$\,Gyr, radius as $0.089^{+0.016}_{-0.011}$\,R$_{\odot}$ and a mass of $0.052^{+0.015}_{-0.022}$\,M$_{\odot}$ are given.
    
    With a scale factor of $(R/d)^2=6.19^{+0.16}_{-0.14}\times10^{-25}$ and the given radius of $0.089^{+0.016}_{-0.011}$\,R$_{\odot}$, we estimated the distance of the source is $2.55^{+0.33}_{-0.48}$\,kpc.
    As the CEERS field points away from the galactic bulge with an angle (l = 96.50\,$^{\circ}$, b = 59.48\,$^{\circ}$), it could possibly be located in the thick disk or galactic halo.
    
\begin{figure}
            \includegraphics[width=\columnwidth]{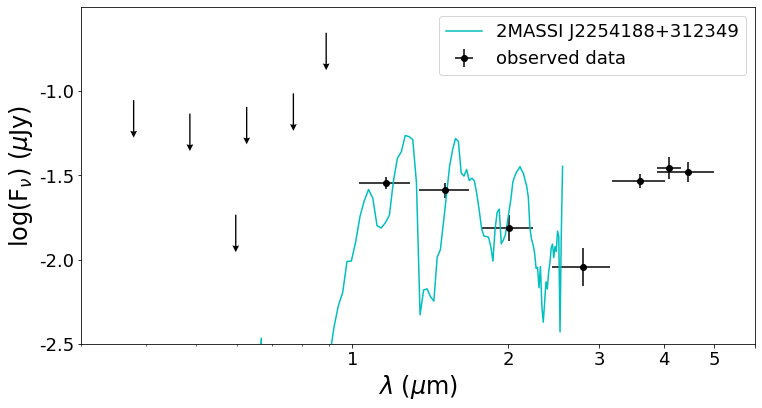}
            \caption{The SED of the T-dwarf candidate. Observed JWST/HST photometry are shown in black points. The cyan line shows the best-fit spectrum from T dwarf IR standard: 2MASSIJ2254188+312349.}
            \label{fig:speccheck}
    \end{figure}

    %Adopting a spectral type to \textit{Spitzer} ch1-ch2 colour relation from \citet[]{2021ApJS..253....7K} Table 13, by using the F356W-F444W=0.14$\pm$0.13 as ch1-ch2, we found CEERS-BD1 is possible to be a L0-T6 object. However, as the temperature of 1300\,K of the object is a transition between L and T spectral type, 
    We compared the \textit{JWST} NIRCam data to L dwarf IR standards \citep{2008AJ....136.1290R} and T dwarf IR standards \citep{2006ApJ...637.1067B} from SpeX library\footnote{\url{https://cass.ucsd.edu/~ajb/browndwarfs/spexprism/library.html}}, and found CEERS-BD1 better match to T4 type as Figure \ref{fig:speccheck} shows. A caveat should be raised here as the standard spectra are ranging from $\sim$1.0 to $\sim$2.4\,$\mu$m, which only covers three filters from \textit{JWST} NIRCam. Therefore we would conclude CEERS-BD1 is possibly an early-T dwarf, where further spectroscopic observation is necessary to infer the precise spectral type.

    \subsection{Space Density}
    
    We also discuss the expected number of early T dwarfs that could be found within the searched fields by adopting the space density from \citet[]{2021ApJS..253....7K}. 
    
    We first calculate the volumes of each observed field. By adopting the detection limit as 5$\sigma$ depth of F444W filter, together with the absolute magnitude of 14.153\,mag. This value is provided by the Sonora photometric table \citep{2021ApJ...920...85M} with parameters of $(T_\text{eff},\log{g},\log{K_\text{zz}}))=(1300, 4.25, 7)$.  
    
    For CEERS observation, adopted from \citet[]{2022arXiv221102495B}, the four fields are having similar F444W limits as 28.57\,mag for CEERS1 field, 28.58\,mag for CEERS2, 28.58\,mag for CEERS3, and 28.58\,mag for CEERS6, which we calculated an approximate search limit of 7.68\,kpc. For SQ and SMACSJ0723, adopted from \citet[]{2022arXiv220801612H}, the F444W limit is 28.3\,mag and 29.6\,mag, corresponding to distances of 6.75\,kpc and 12.3\,kpc respectively. According to \citet[]{2022ApJ...938L..14M}, the 5$\sigma$ depth of F444W is 29.71\,mag for GLASS field, which would relate to a distance of 12.9\,kpc. The field of view for each field is calculated directly from the image, with 9.68\,arcmin$^2$ for each CEERS field and GLASS,  43.2\,arcmin$^2$ for SQ, and 10.1\,arcmin$^2$ for SMACSJ0723.
    
    Considering the scale height of the galactic thin disk as 300\,pc and the pointing direction of each field  (CEERS as $l,b=(96.50^{\circ}, 59.48^{\circ})$, SQ as $l,b=(93.26^{\circ}, -20.99^{\circ})$, SMACSJ0723 as $l,b=(285.01^{\circ}, -23.74^{\circ})$, GLASS as $l,b=(9.52^{\circ}, -81.19^{\circ})$), with adopting the space density of 1.95$\pm$0.30 ($\times10^{-3}\text{pc}^{-3}$) for 1200\,K-1350\,K objects from \citet[]{2021ApJS..253....7K} Table 15, we derived the expected observed number of such objects in each field as follows: CEERS as $(50.3\pm 7.7)\times10^{-3}$ for each, SQ as $(17.7\pm 2.7)\times10^{-3}$, SMACSJ0723 as $(5.89\pm 0.91)\times10^{-3}$, and GLASS as $(83.3\pm 12.8)\times10^{-3}$.  
    
    Although the expected numbers are small, it would not be impossible. As the census from \citet[]{2021ApJS..253....7K} was complete through 20\,pc, the space density for thick disk/ halo population may be different where the \textit{JWST} NIRCam search volume is mostly beyond thin disk. A recent study \citep{2022ApJ...924..114A} showed numbers of L and T-dwarfs with a distance of $\sim$2kpc and $\sim$400pc, could be discovered from \textit{Hubble Space Telescope} Wide Field Camera 3 observations,  but more distant brown dwarfs need JWST such as in this work. We clearly need a larger volume to accurately measure the space density of brown dwarfs. Such future JWST observations are awaited.
    
\section{Conclusions}%_______________________________________
\label{sec:conclusion}
    
    This work aims to select brown dwarf candidates from the ERO and ERS of \textit{JWST} using NIRCam data. With specific colour selection criteria of F115W$-$F277W$<$-0.8 and F277W$-$F444W$>$1.1, we present a discovery of a brown dwarf candidate from the \textit{JWST} CEERS field. 
    
    A best-fitted theoretical SED model suggested that CEERS-BD1 is an early T-dwarf with the effective temperature of T$_\text{eff}\approx$1300\,K. An estimated distance of $\approx2.55$\,kpc shows it could possibly be located in the thick disk or galactic halo. The follow-up CEERS MIRI observation covering the position of CEERS-BD1 would further provide information to this source. Also spectroscopic observations such as \textit{JWST} NIRSpec would be important to further confirm the properties of this source.

\section*{Acknowledgements}%_________________________________
\label{sec:acknowledgement}
The authors would like to appreciate the constructive suggestions and comments from the anonymous referee which greatly improved the quality of this manuscript.
TG acknowledges the support of the National Science and Technology Council of Taiwan through grants 108-2628-M-007-004-MY3 and 110-2112-M-005-013-MY3. TH acknowledges the support of the National Science and Technology Council of Taiwan through grants 110-2112-M-005-013-MY3, 110-2112-M-007-034-, and 111-2123-M-001-008-.

\section*{Data Availability}%________________________________
\textit{JWST}/NIRCam images used in this work are available through the Mikulski Archive for Space Telescopes (\href{https://mast.stsci.edu/}{https://mast.stsci.edu/}). Additional data products and analysis code will be made available upon reasonable request to the corresponding author.

\label{sec:data_availablility}


\bibliographystyle{mnras}
\bibliography{paper} % if your bibtex file is called example.bib

%%%%%%%%%%%%%%%%%%%%%%%%%%%%%%%%%%%%%%%%%%%%%%%%%%

%%%%%%%%%%%%%%%%% APPENDICES %%%%%%%%%%%%%%%%%%%%%

%%%%%%%%%%%%%%%%%%%%%%%%%%%%%%%%%%%%%%%%%%%%%%%%%%

% Don't change these lines
\bsp	% typesetting comment
\label{lastpage}
\end{document}